\documentclass{MRF}

\usepackage{epstopdf}
\usepackage{hyperref}
\usepackage{xcolor}
\usepackage{graphicx}
\usepackage{subcaption}
\usepackage{siunitx}
\usepackage{booktabs}
\usepackage{multirow}
\usepackage[table]{xcolor} 
\usepackage{comment}
\usepackage{balance}

\begin{document}

\supertitle{Research Paper}

\title[Millimeter-Wave UAV Channel Model]{
Millimeter-Wave UAV Channel Model with Height-Dependent Path Loss and Shadowing in Urban Scenarios}

\author[Saboor et al.]{Abdul Saboor$^{1}$ and Evgenii Vinogradov$^{1,2}$}

\address{
\add{1}{WaveCoRE - Department of Electrical Engineering (ESAT), KU Leuven, Belgium}
\add{2}{NaNoNetworking Center in Catalonia (N3Cat), Universitat Polit\`{e}cnica de Catalunya, Spain}
}

\corres{\name{Abdul Saboor}
\email{abdul.saboor059@gmail.com}}

\begin{abstract}

Uncrewed Aerial Vehicles (UAVs) serving as Aerial Base Stations (ABSs) are expected to extend 6G millimeter-Wave (mmWave) coverage and improve link reliability in urban areas. However, UAV-based Air-to-Ground (A2G) channels are highly dependent on height and urban geometry. This paper proposes an ABS height-dependent mmWave channel model and investigates whether urban geometry, beyond the standard built-up parameters, significantly affects LoS probability ($P_{\text{LoS}}$) and Large-Scale Fading (LSF). Using MATLAB ray tracing at 26 GHz, we simulate approximately 10K city realizations for four urban layouts that share identical built-up parameters but differ in their spatial organization. We extract elevation-based $P_{\text{LoS}}$ using a sigmoid model and derive height-dependent Path-Loss Exponents (PLEs) and shadow-fading trends using exponential fits. Results show that PLE for Non-Line-of-Sight (NLoS) decreases toward $2.5$--$3$ at high altitudes, Line-of-Sight (LoS) PLE remains near $2$, and shadow fading reduces with height. We also find that geometric layout introduces a modest but consistent change in PLE ($\approx \pm 0.2$), even when built-up parameters are fixed. The proposed model matches ray-tracing results and provides a practical, height-dependent LSF model for ABS planning in urban scenarios.

\end{abstract}

\maketitle

\section{Introduction} 
Uncrewed Aerial Vehicles (UAVs) are emerging as key enablers of future 6G networks~\cite{othman2025key}. UAVs carrying communication payloads, known as Aerial Base Stations (ABSs), can be deployed to extend coverage and address temporary connectivity needs in congested or limited-access areas~\cite{saboorTVT}. The key benefit of ABS over a Terrestrial Base Station (TBS) is its 3D movement, enabling Line-of-Sight (LoS) maximization and improved coverage~\cite{chaalal2022new}, making it well-suited for dense urban environments.

However, Air-to-Ground (A2G) links behave very differently from traditional terrestrial links. In particular, urban geometric features like building density, height, street layout, and spatial randomness have a significant impact on A2G signal propagation~\cite{pang2022geometry}. These geometric factors strongly influence the LoS availability, Path Loss (PL), and Large-Scale Fading (LSF) statistics~\cite{AbdulAWPL}. As UAVs move across different altitudes, the channel transitions between LoS and Non-Line-of-Sight (NLoS), leading to altitude-dependent path loss that terrestrial models cannot capture~\cite{saboor2025cash}. Similar height-dependent behavior has also been observed in sub-6~GHz field measurements, where Path Loss Exponent (PLE) decreases toward free-space conditions as UAV altitude increases~\cite{7936620}. These effects become even more prominent at millimeter-Wave (mmWave) frequencies, where signals are highly sensitive to blockage and geometric variations~\cite{ma2025effects}. Therefore, accurately modeling this transition is essential for designing reliable ABS deployment strategies.

Several analytical A2G channel models have been proposed in the literature to model the LoS probability $P_{\text{LoS}}$ and PL in urban environments~\cite{AbdulAWPL,pang2022geometry, ITU, Saboor2023plos, hourani2014, AbdulEucap}. Recent survey studies have highlighted rapid progress in UAV channel modeling and channel sounding, emphasizing the need for measurement-driven and comparative modeling frameworks~\cite{seah2024empirical,10616106}.
A widely adopted $P_{\text{LoS}}$ model is provided by the International Telecommunication Union (ITU), developed for simplified Manhattan-type environments with uniformly distributed buildings characterized by parameters such as building density and height~\cite{ITU}. Because of this simplicity, several other studies used Manhattan to model $P_{\text{LoS}}$ and PL~\cite{AbdulAWPL, hourani2014, Saboor2023plos}. However, real urban environments rarely exhibit such regularity; building shapes, orientations, and street layouts vary significantly, and these geometric variations strongly affect $P_{\text{LoS}}$ and LSF. This limitation becomes more significant at mmWave frequencies. In urban mmWave measurements, Akdeniz et al.~\cite{6834753} highlight sharp LoS/NLoS transitions where the channel is extremely sensitive to the surrounding geometry. In contrast, RT studies of mmWave A2G links reveal altitude-dependent two-ray behavior in open areas and severe fluctuations in dense, high-rise regions \cite{8288376}. More advanced UAV-focused mmWave models further account for 3D antenna patterns, side-lobe interference, UAV angular vibrations, and orientation misalignment, all of which significantly influence A2G path loss and fading \cite{7936620, 9904477}. Recent neural-network models further demonstrate that mmWave A2G channels contain rich geometric dependencies that simplified stochastic or 3rd Generation Partnership Project (3GPP)-style models fail to capture~\cite{9782737}.

This leads to a fundamental unresolved issue: \emph{To what extent does building geometry, beyond the standard built-up parameters, influence $P_{\text{LoS}}$ and the resulting LSF in A2G channels?} Models in the existing literature implicitly assume that environments with identical built-up parameters (i.e., the number of buildings, their density, and height) will exhibit similar LSF behavior, regardless of their spatial layout. However, this assumption has not been systematically validated for A2G links, particularly at altitudes where spatial layout may significantly influence propagation behavior.

This paper addresses this gap by examining how geometric randomness affects LSF when the underlying statistical built-up parameters are fixed. Using MATLAB Ray-Tracing (RT), we quantify how altitude-dependent PLE and shadow fading vary across these layouts. Our findings help determine whether A2G LSF is driven primarily by geometry or by built-up statistics alone. \textbf{\textit{An earlier version of this work was presented at the 2025 European Conference on Antennas and Propagation (EuCAP) and published in its Proceedings~\cite{AbdulEucap}. This article substantially extends the conference version by introducing height-dependent LSF analysis across four urban environments and different layouts.}} The main contributions and findings of this work are summarized as follows:

\begin{itemize}
    \item We investigate whether urban geometry, beyond standard ITU built-up parameters $(\alpha,\beta,\gamma)$, affects A2G channel statistics. Results show that spatial layout has a limited impact on $P_{\text{LoS}}$, but slight changes in height-dependent LSF, particularly under NLoS, with consistent PLE variations of $\pm 0.2$ across layouts.

    \item We show that the NLoS PLE follows a consistent height-dependent trend across all environments and layouts, approaching approximately $2.5$--$3$ at high altitudes, while LoS remains close to free space.

    \item We propose a compact unified LSF model for mmWave A2G channels capturing elevation-dependent $P_{\text{LoS}}$, PLE, and shadow fading for regular and randomized layouts.

    \item We validate the model via Kullback--Leibler divergence, showing it reproduces RT statistics across all scenarios. 
\end{itemize}

The rest of the paper is organized as follows. Section~\ref{sec2} presents the system model and the proposed urban layouts. Section~\ref{sec3} details the LoS and LSF parameterization procedures. Section~\ref{sec4} analyzes the simulation results and discusses the impact of layout randomness on A2G propagation. Finally, Section~\ref{sec5} concludes the paper.

\section{System model} 
\label{sec2}
This section describes different urban layouts considered for A2G channel modeling in this paper. We begin by outlining the standard ITU-based Manhattan grid, which serves as the baseline model, and then introduce three generalized layouts to capture irregular city structures.

\subsection{Standard ITU Manhattan layout}

\begin{figure*}[!t]
   \centering
    \begin{subfigure}[b]{0.35\linewidth}
    \includegraphics[width=\linewidth]{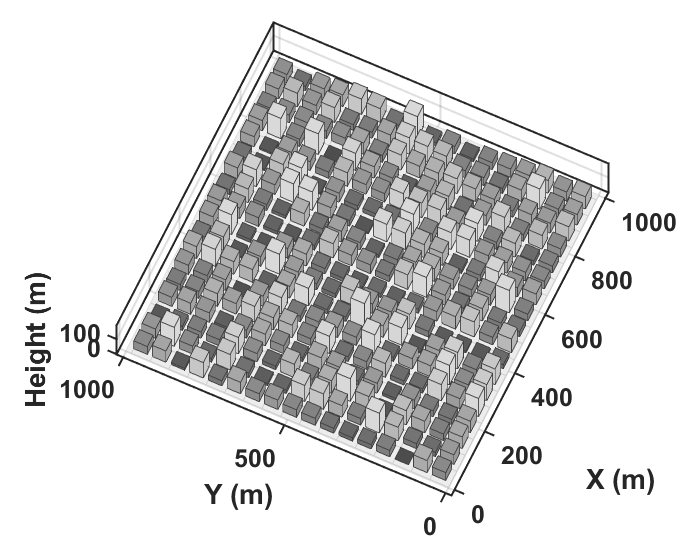}
    \caption{Manhattan Layout}
    \label{simManhat}
  \end{subfigure}
  \centering
    \begin{subfigure}[b]{0.35\linewidth}
    \includegraphics[width=\linewidth]{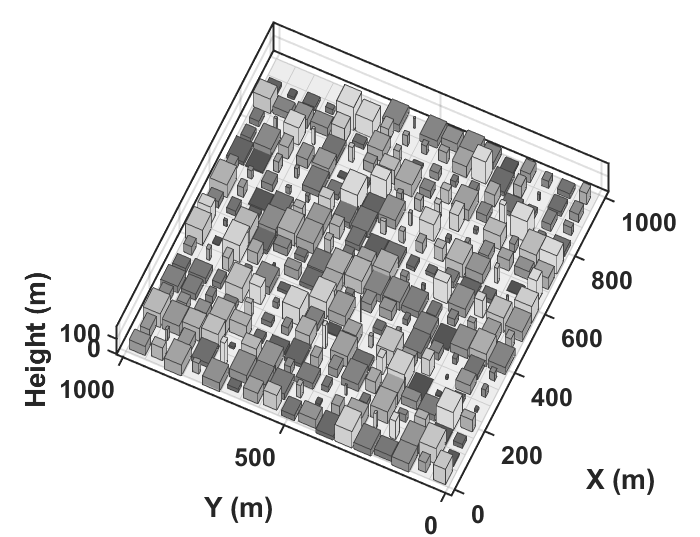}
    \caption{SRU Layout}
    \label{sim1}
  \end{subfigure}
  \centering
  \begin{subfigure}[b]{0.35\linewidth}
    \includegraphics[width=\linewidth]{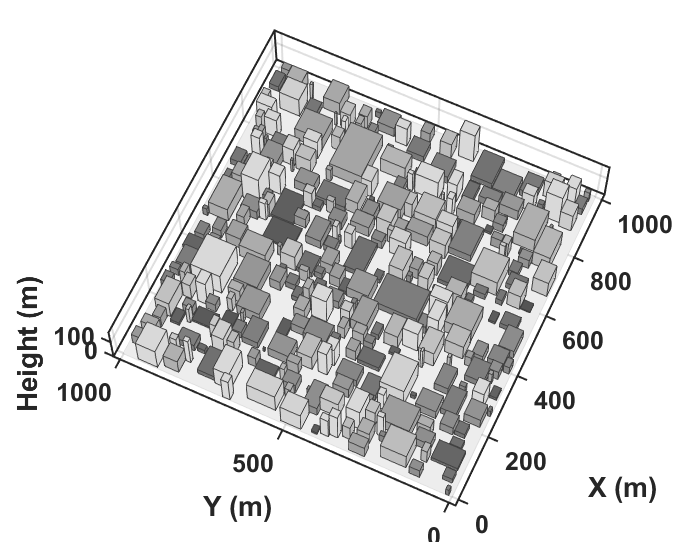}
    \caption{FUU Layout}
    \label{sim2}
  \end{subfigure}
  \centering
  \begin{subfigure}[b]{0.35\linewidth}
    \includegraphics[width=\linewidth]{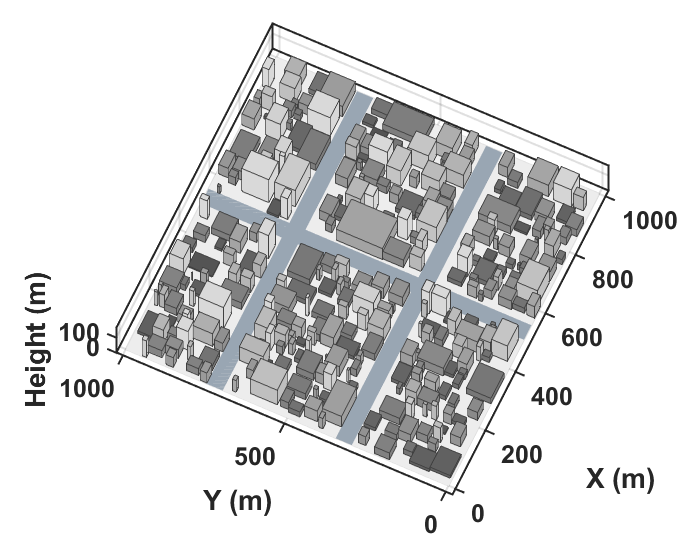}
    \caption{HEU Layout}
    \label{sim3}
  \end{subfigure}
  \caption{Manhattan and proposed urban layouts for high-rise urban environment ($\alpha = 0.5, \beta = 300, \gamma = 50$).} 
  \label{fig4}
  \vspace{-1em}
\end{figure*}

ITU characterizes urban environments through a tuple of built-up parameters $(\alpha, \beta, \gamma)$, where $\alpha$ is the fraction of built area, $\beta$ represents the number of buildings per km$^2$, and $\gamma$ is the Rayleigh scale parameter for building height~\cite{ITU}. Based on these parameters, different urban layouts can be generated~\cite{ITU} within a Manhattan grid by arranging identical square buildings separated by evenly spaced streets, as shown in Fig.~\ref{simManhat}. The building width $W$ and street width $S$ are derived as $W = 1000\sqrt{\frac{\alpha}{\beta}}, S = 1000\sqrt{\frac{1}{\beta}} - W$, within a $1$~km$^2$ reference area. Building heights follow the Rayleigh probability density function
\begin{equation}
\label{rayleigh}
f(h) = \frac{h}{\gamma^2}e^{-h^2/(2\gamma^2)},
\end{equation}

The simplicity and geometry of the Manhattan layout make it a popular reference for theoretical analysis, such as $P_{\text{LoS}}$ and $PL$. However, it fails to represent the geometric diversity of real cities. These limitations motivate the development of more flexible and irregular urban layouts presented in the following subsections.

\subsection{Proposed urban layouts}

To investigate whether geometric randomness in city structures influences LSF when the underlying built-up parameters are the same, we develop three random urban layouts using identical $(\alpha, \beta, \gamma)$ to ensure consistent building density, coverage, and height distribution~\cite{saboorWCNC}. The only distinguishing factor among the layouts is the organization of buildings and open spaces in the total area $A_\text{total}$. The total area occupied by buildings is given by $A_\beta = \alpha A_\text{total}$ and $B_{\text{avg}} = \frac{A_\beta}{\beta}$.

\subsubsection{Structured Random Urban (SRU) layout}

The SRU layout maintains the general Manhattan alignment but introduces variability in building geometry/size and spacing, as visualized in Fig.~\ref{sim1}. This controlled irregularity allows investigation of how moderate geometric randomness affects LSF compared to a perfectly regular grid. Each city block is divided into $\beta$ sections, representing the number of buildings per square kilometer. The area of the $i$-th building is then randomly varied around the average building area $B_{\text{avg}}$ as
\begin{equation}
A_i = B_{\text{avg}}(0.6 + 0.8r_i), \qquad r_i \sim \mathcal{U}(0,1),
\end{equation}
ensuring that $A_i \in [0.6B_{\text{avg}}, 1.4B_{\text{avg}}]$. 
For each building, the width $W_i$ and length $L_i$ are determined as
\begin{equation}
W_i = \sqrt{A_i}(0.5 + r_i), \qquad L_i = \frac{A_i}{W_i},
\end{equation}
while the height follows the Rayleigh distribution from equation \eqref{rayleigh}.

Buildings are placed in a grid, creating structured streets with irregular building areas. This hybrid layout represents a more realistic downtown area where city grids exist, yet individual buildings differ in size and shape. As a result, the SRU acts as an intermediate case between deterministic and entirely random city structures, allowing a controlled analysis of how geometric variability affects the LSF.

\subsubsection{Fully Unstructured Urban (FUU) layout}
The FUU layout represents the extreme case of geometric randomness, where no predefined grid or symmetry is considered. In FUU, buildings are distributed freely across the simulation area, creating highly irregular streets and non-uniform building clusters, as illustrated in Fig.~\ref{sim2}. 
This layout resembles old cities, where building placement and spacing are often highly irregular. Given the total building area $A_\beta$, the individual building areas are assigned using a Dirichlet distribution
\begin{equation}
\mathbf{A} = [A_1, A_2, \ldots, A_\beta] \sim \text{Dirichlet}(\mathbf{1}_\beta)\,A_\beta,
\end{equation}
which ensures that $\sum_i A_i = A_\beta$. Each building is then assigned a random position $(x_i, y_i)$ within the $A_\text{total}$ region, with an overlap constraint defined as
\begin{equation}
G(x,y) = 
\begin{cases}
1, & \text{if occupied by a building,}\\
0, & \text{otherwise,}
\end{cases}
\end{equation}
\begin{equation*}
G(x_i:x_i+W_i,\,y_i:y_i+L_i) = 0.
\end{equation*}

The resulting layout provides maximum randomness in spatial geometry while maintaining identical statistical density and height characteristics. This allows evaluation of whether LSF behavior depends primarily on environmental density parameters $(\alpha, \beta, \gamma)$ or on the spatial organization of the same built-up area.

\subsubsection{Highway-Embedded Urban (HEU) layout}
The HEU layout is an extension of the FUU layout where we embed $n_H$ linear corridors representing highways in the environment, as illustrated in Fig.~\ref{sim3}. These highways create open spaces within dense areas, mimicking mixed urban environments composed of large highways and compact residential blocks around them. HEU helps in examining how large open corridors affect overall propagation and path-loss behavior.

Let $n_H$ denote the number of highways, each defined by its width $W_H^{(j)}$, length $L_H^{(j)}$, and orientation $\phi_H^{(j)} \in [0, \pi)$. The total area occupied by highways is given by
\begin{equation}
A_H = \sum_{j=1}^{n_H} W_H^{(j)} L_H^{(j)}.
\end{equation}
This area is subtracted from the available built-up region before placing buildings, resulting in $A'_\beta = A_\beta - A_H$, where $A'_\beta$ is the effective built-up area after reserving highway space. No buildings are placed inside highway regions, enforced by the spatial constraint
\begin{equation}
G(x,y) = 0, \qquad \forall (x,y) \in \Omega_H,
\end{equation}
where $\Omega_H$ represents the set of grid cells covered by highways.

Buildings are then distributed in the remaining area using the same procedure used in the FUU layout. This combination of dense building clusters and highways helps in examining how the presence of wide corridors within urban areas influences LSF and the resulting PLE.

\vspace{-.2cm}
\subsection{Air-to-ground Channel Model}
\label{subsec:channel_model}
The A2G channel between the UAV/ABS and a ground user can be represented as a combination of deterministic free-space and random losses due to urban obstructions. The average large-scale attenuation $\Lambda(d, h_{\text{ABS}})$ depends on both the 3D distance $d$ and the ABS height $h_{\text{ABS}}$ (or simply on elevation $\theta$), which can be expressed as a weighted sum of LoS and NLoS components
\begin{equation}
\label{eq:channel_model}
\begin{aligned}
\Lambda(d,h_{\text{ABS}}) =\;& P_{\text{LoS}}(\theta)\,\Lambda_{\text{LoS}}(d,h_{\text{ABS}}) \\
&+ \bigl[1-P_{\text{LoS}}(\theta)\bigr]\Lambda_{\text{NLoS}}(d,h_{\text{ABS}}).
\end{aligned}
\end{equation}
Here, $P_{\text{LoS}}(\theta)$ depends on the elevation angle $\theta = \tan^{-1}\!\bigl((h_{\text{ABS}} - h_{\text{GU}})/r\bigr)$, 
which is linked to the 3D distance $d = \sqrt{r^2 + (h_{\text{ABS}} - h_{\text{GU}})^2}$, and $h_{\text{GU}}$ is GU height. 

This formulation provides a unified representation of the A2G channel by combining distance-dependent attenuation with the probabilistic influence of the surrounding geometry. The following subsection details $P_{\text{LoS}}$ modeling for different urban configurations.

\subsubsection{LoS Probability Modeling}

Ideally, $P_{\text{LoS}}$ depends on both the elevation $\theta$ and azimuth $\varphi$ angles, which can be computed using the geometry-based formulation presented in~\cite{Saboor2023plos}. However, since the azimuthal orientation of streets and GU positions tends to vary randomly in realistic city layouts, its effect averages out statistically over multiple realizations. Therefore, the LoS probability is typically modeled as an elevation-only function $P_{\text{LoS}}(\theta)$.

In our recent work~\cite{saboorWCNC}, we demonstrated that the empirical $P_{\text{LoS}}$ can be accurately represented by a four-parameter Sigmoid model as
\begin{equation}
\label{eq:sigmoid}
P_{\text{LoS}}(\theta) = \frac{1}{1 + \exp(x_1 \theta^3 + x_2 \theta^2 + x_3 \theta + x_4)},
\end{equation}
where $x_1, x_2, x_3, x_4$ are environment-dependent coefficients obtained via nonlinear curve fitting from large-scale RT data.

\subsubsection{Large-Scale Fading Modeling}
\vspace{.5em}
\noindent
\textbf{Impact of Buildings:} Buildings are the dominant source of LSF in urban environments, as they directly impact LoS condition. Accordingly, the LSF is modeled as a LoS-dependent log-distance PL with shadow fading
\begin{equation}
\Lambda_{x}(d,h_\text{ABS}) =
\Lambda_0 + 10\, n_x(h_\text{ABS}) \log_{10}(\frac{d}{d_0}) + \Psi_x,
\end{equation}
where $x$ can be LoS or NLoS, $\Lambda_0$ is the reference free-space path loss at distance $d_0=1$~m, $n_x(h_\text{ABS})$ are height-dependent PLEs, and $\Psi_x$ is the zero-mean Gaussian shadow fading term with height-dependent standard deviations $\sigma_x(h_\text{ABS})$. 

As the ABS height increases, the number of potential obstructions decreases and the elevation angle improves, leading to a gradual transition from NLoS to LoS dominance. We model the height-dependent PLE as
\begin{equation}\label{eq:PLE_fit}
n(h_\text{ABS}) = n_\infty + (n_0 - n_\infty) \exp(-h_\text{ABS}/h_0^n),
\end{equation}
where $n_0$ and $n_\infty$ are limiting PLEs and $h_0$ is a characteristic decay height. The height-dependent standard deviation of shadow fading is similarly modeled as
\begin{equation}\label{eq:std_fit}
\sigma(h_\text{ABS}) = \sigma_\infty + (\sigma_0 - \sigma_\infty) \exp(-h_\text{ABS}/h_0^\sigma),
\end{equation}
with $\sigma_0$ and $\sigma_\infty$ serving as limiting values.

\vspace{.5em}
\noindent
\textbf{Impact of Random Obstructions:} The proposed model focuses on LSF caused by buildings, which dominate attenuation in urban A2G links. In practice, smaller obstacles such as trees or streetlights may introduce additional losses, especially at mmWave frequencies. However, as shown in our earlier work~\cite{AbdulEucap}, the excess loss caused by such objects is typically distance-independent and remains below 3~dB on average, leaving the height-dependent PLE largely unaffected. Therefore, these effects are not explicitly modeled here but can be incorporated as a fixed attenuation offset when finer accuracy is required.

\vspace{-.3cm}
\section{Model Parameterization}
\label{sec3}

This section describes how $P_{\text{LoS}}$ and LSF parameters are obtained from RT data for the considered urban layouts. The simulations were performed using a MATLAB-based Shooting and Bouncing Rays (SBR) engine to generate spatially consistent datasets of PL, shadow fading, and LoS/NLoS states over a $1000 \times 1000~\mathrm{m^2}$ area with $20~\mathrm{m}$ RX grid resolution. The ABS height varies from $5$ to $1000~\mathrm{m}$ (30 heights), with ground receivers at $1.5~\mathrm{m}$. Each configuration is evaluated over 20 independent city realizations across four layouts (Manhattan, SRU, FUU, and HEU) and four environments (suburban to high-rise urban). The RT simulator considers up to 3 reflections and 1 diffraction per path at $26~\mathrm{GHz}$ with $46~\mathrm{dBm}$ transmit power and isotropic antennas. Isotropic antennas are adopted to characterize the intrinsic propagation behavior of the environment, independent of antenna directivity, which can be incorporated separately in system-level analysis. Furthermore, diffuse scattering is not explicitly modeled. Although present at mmWave frequencies, it mainly affects small-scale channel characteristics. Since this work focuses on LSF, the dominant effects are captured by LoS propagation, specular reflections, and diffraction. Including diffuse scattering would also significantly increase computational complexity without materially affecting the extracted PLE and shadow fading trends. The complete set of simulation parameters is summarized in Table~\ref{rt_params}. 

\begin{table}[!t]
\centering
\small
\caption{Ray-Tracing simulation parameters and configuration}
\label{rt_params}
\begin{tabular}{ll}
\hline
\textbf{Parameter} & \textbf{Value / Model} \\
\hline
\textbf{System \& Engine Setup} & \\
Carrier Frequency & 26~GHz \\
Transmit Power & 46~dBm \\
Antenna Type (Tx \& Rx) & Isotropic (0~dBi) \\
RT Engine & MATLAB SBR \\
Allowed Interactions & 3 reflections and 1 diffraction \\
Scattering Model & Specular + UTD diffraction \\
\hline
\textbf{Geometry \& Sampling} & \\
City Domain Area & $1000 \times 1000~\mathrm{m^2}$ \\
Rx Grid Resolution & 20~m \\
Ground Receiver Height & 1.5~m \\
ABS (UAV) Heights & 5-1000~m (30 discrete heights) \\
City Realizations & 20 per height \\
Total Samples Evaluated & Up to 50,000 per height \\
\hline
\end{tabular}
\end{table}

\subsection{LoS Parameterization}
\noindent
\textbf{Parameter extraction:} 
In the RT simulations, $P_{\text{LoS}}$ or LoS condition is derived directly from geometric visibility between the ABS and GU. For each ABS–GU pair, the RT checks whether the straight line between them intersects any building. If there is no intersection, the link is LoS; else, it is NLoS. For each ABS height, multiple city realizations and user locations are simulated to capture statistical variability. The empirical $P_{\text{LoS}}(\theta)$ is obtained as $P_{\text{LoS}}(\theta) = \frac{N_{\text{LoS}}(\theta)}{N_{\text{Total}}(\theta)}$, where, $N_{\text{LoS}}(\theta)$ and $N_{\text{Total}}(\theta)$  are the number of LoS and total links observed for a particular $\theta$, respectively. 
Therefore, $P_{\text{LoS}}$ is estimated by fitting~\eqref{eq:sigmoid} across urban environments and layouts.

\noindent
\textbf{Extracted parameters:} 
The fitting parameters for the four standard urban environments, corresponding to each layout, are given in Table~\ref{tab:model}. In addition to the four layouts described in Section~\ref{sec2}, we provide the generalized results based on the combined data of all the layouts.

\subsection{LSF Parameterization}
\noindent
\textbf{Parameter extraction:} 
LSF statistics are obtained from the RT dataset by separating LoS/NLoS links and estimating height-dependent PLEs and shadow fading distributions. From the RT dataset, LoS and NLoS channels are separated for each UAV height (all 20 city realizations are used). PLE is obtained via linear regression
\begin{equation}
n(h_\text{ABS})  = \frac{\Lambda_\text{RT}(h_\text{ABS})  - \Lambda_0}{10 \log_{10}(d/d_0)},
\end{equation}
and shadow fading as deviations from the fitted path loss
\begin{equation}
\Psi (h_\text{ABS}) = \Lambda_\text{RT}(h_\text{ABS})  - (\Lambda_0 + 10 n(h_\text{ABS}) \log_{10}(\frac{d}{d_0})).
\end{equation}
Height-dependent PLEs and shadow fading standard deviations are fitted with exponential functions \eqref{eq:PLE_fit} and \eqref{eq:std_fit}. 

\noindent
\textbf{Extracted channel parameters:} The extracted parameters are summarized in Table~\ref{tab:params_full}. Characteristic parameters $n_0$, $n_\infty$, $\sigma_0$, $\sigma_\infty$, and $h_0$ provide physical insight into propagation near the ground and asymptotic high-altitude behavior with lower PLE values. 

\begin{figure}[t]
    \centering
    \includegraphics[width=1\linewidth]{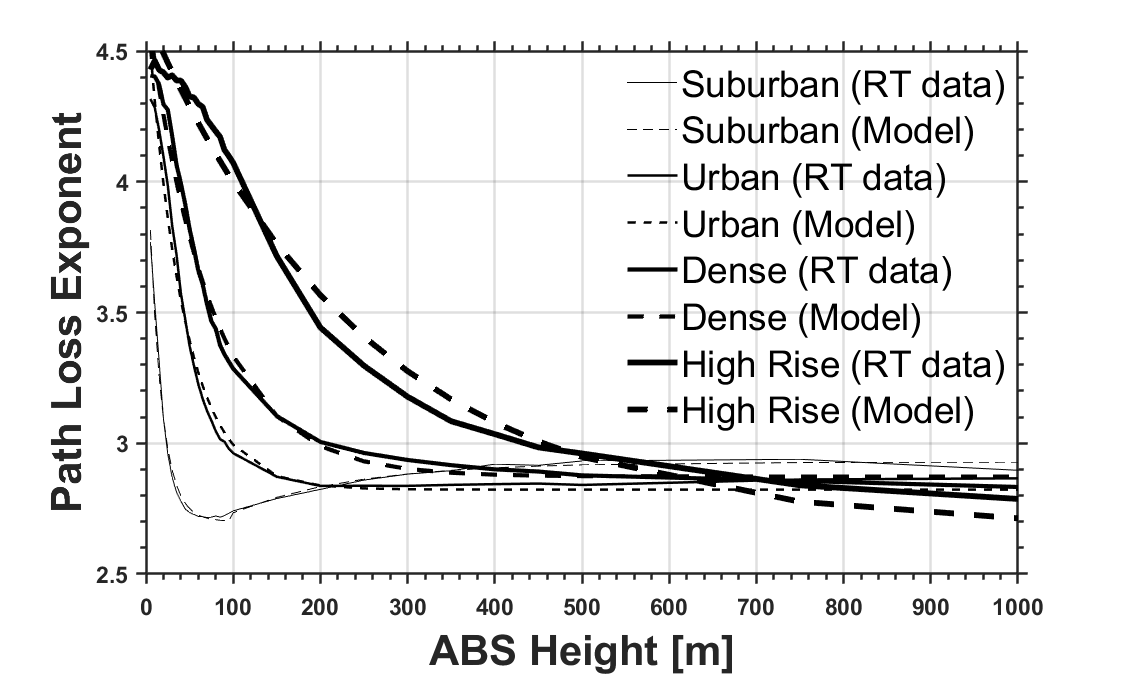}

    \caption{Combined Layouts PLE across ABS altitude and environment under NLoS conditions.}
    \label{fig:ple_model}
    \vspace{-.4cm}
\end{figure}

The results of PLE modeling for the combined data illustrated in Fig.~\ref{fig:ple_model} reveal clear trends in ABS heights and layouts for NLoS conditions. The results reveal that the height-dependent PLEs decrease toward the $2.5-3$ limit at high ABS altitudes, reflecting the gradual transition from heavily obstructed to more favorable propagation. Suburban scenarios display a two-regime behavior, with a breaking point around $100$~m separating a descending PLE regime at low altitudes from an ascending regime at higher altitudes. In contrast, urban and high-rise layouts follow a single descending PLE regime. Note that the LoS PLE equals 2 (free space) in all cases. Shadow fading standard deviations decrease with height, with LoS channels exhibiting lower variability than NLoS links, as expected.

\begin{figure*}[t]
    \centering

    \begin{subfigure}[t]{1\columnwidth}
        \centering
        \includegraphics[width=\linewidth]{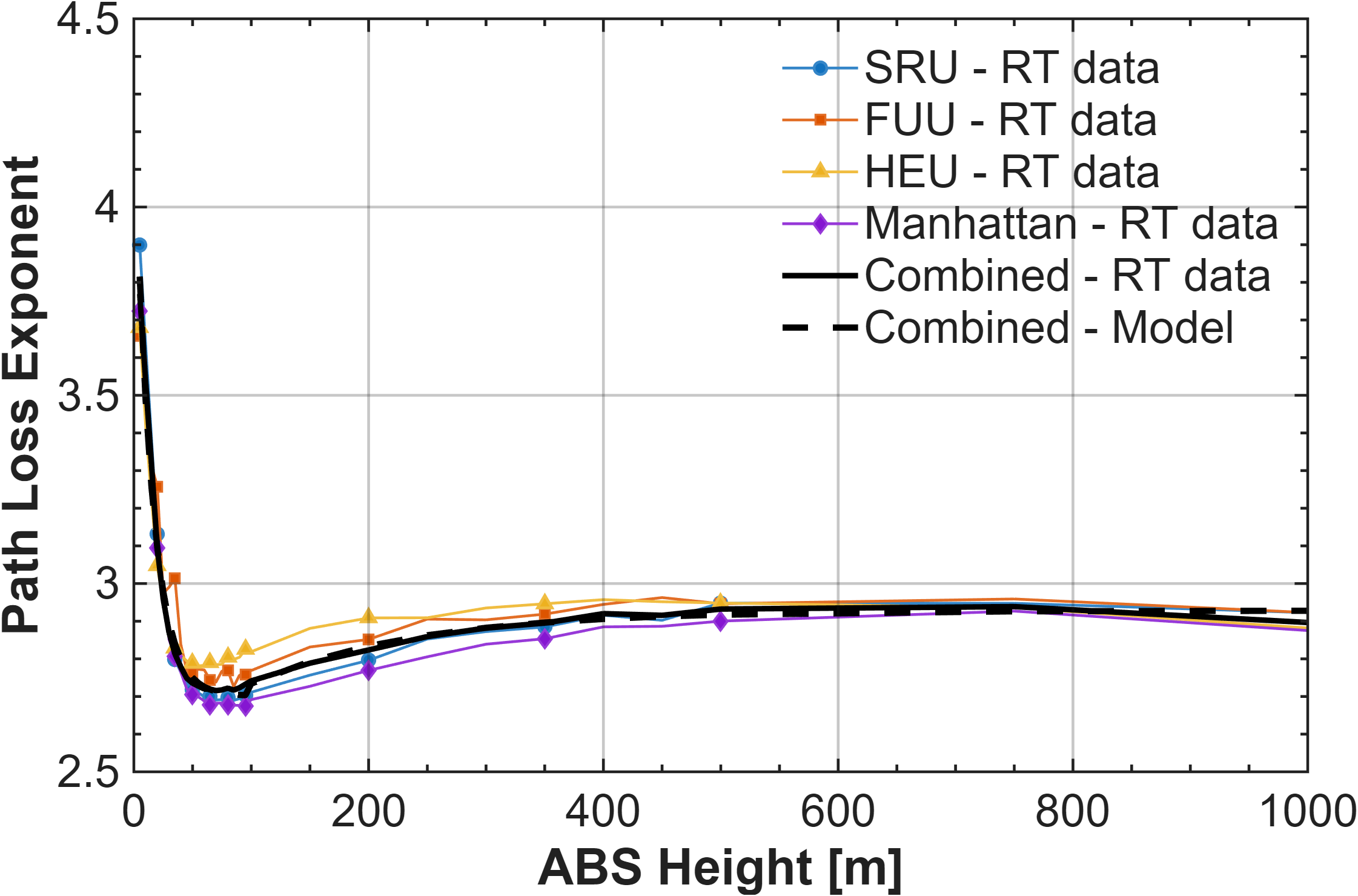}
        \caption{Suburban}
        \label{fig:ple_suburban}
    \end{subfigure}
    \hfill
    \begin{subfigure}[t]{1\columnwidth}
        \centering
        \includegraphics[width=\linewidth]{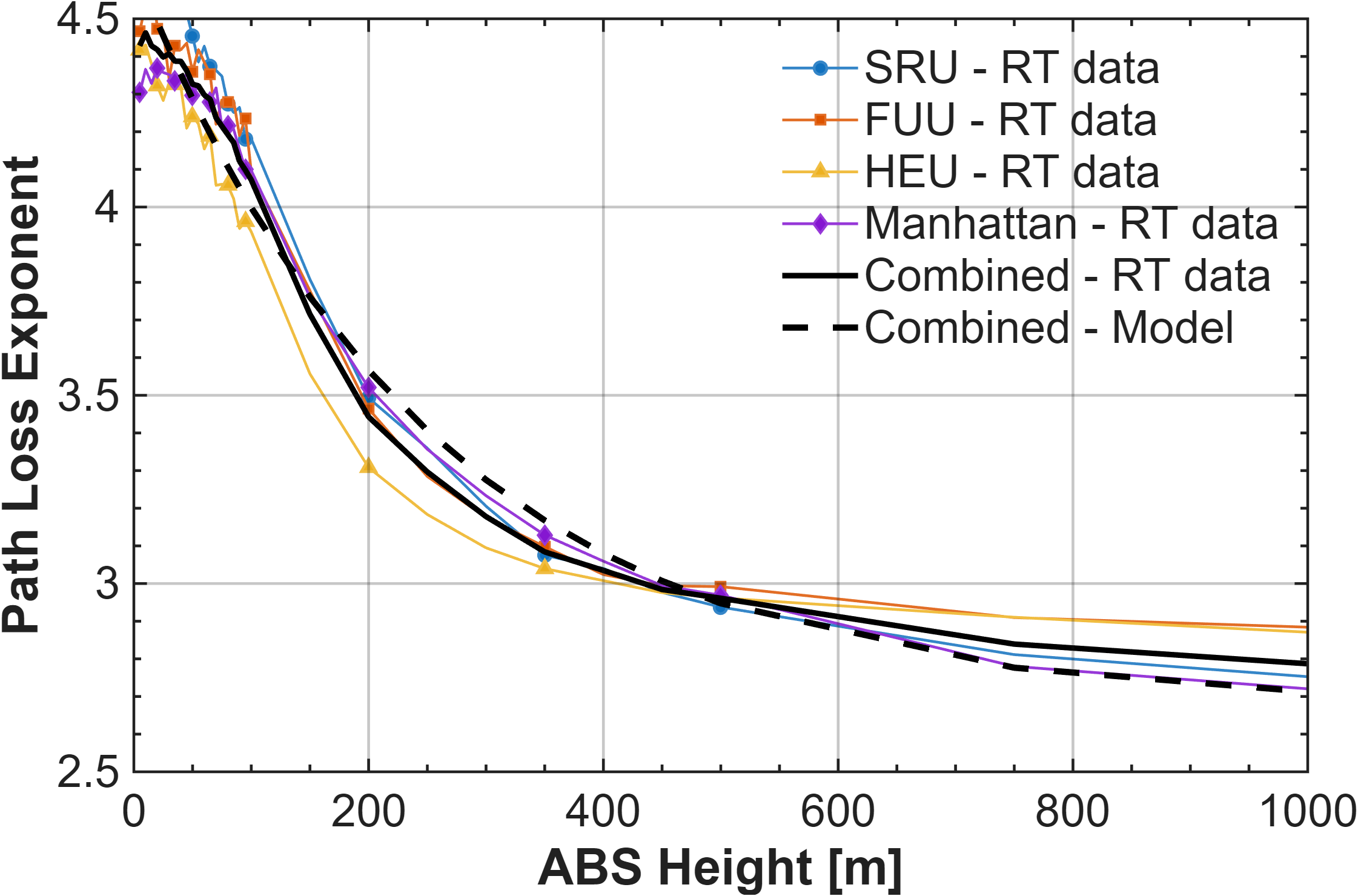}
        \caption{High-rise Urban}
        \label{fig:ple_HR}
    \end{subfigure}
        \caption{NLoS PLE as a function of ABS deployment altitude for the four urban environments, the combined data, and the fitted height-dependent model for the combined data.}
    \label{fig:PLE_vs_height}
\end{figure*}

More detailed PLE results are provided in Fig.~\ref{fig:PLE_vs_height} for all four layouts and the combined data. It can be observed that the layout has a relatively moderate effect ($\pm$0.2) on the PLE behavior. Since such a difference may still be significant in some cases, Table~\ref{tab:model} contains all the details. However, in the further analysis, we mostly focus on the combined data, as it represents a general layout.

The fitted exponential models achieve Root Mean Square Errors (RMSEs) of $0.02$--$0.11$ (PLE) and $0.07$--$0.36$ (LoS) and $0.31$--$0.95$ (NLoS) for shadow fading standard deviation, indicating good agreement with the RT data. 

\begin{table*}[!t]
\centering
\small
\renewcommand{\arraystretch}{1.05}
\caption{Extracted model parameters for LoS and NLoS links across all layouts and height ranges}\label{tab:model}
\begin{tabular}{|l||cccc||cccc||cccccc||c|}
\hline
\multirow{2}{*}{\textbf{Environment}}  & \multicolumn{4}{c||}{\textbf{LoS probability model}}&\multicolumn{4}{c||}{\textbf{LoS channel model}} & \multicolumn{6}{c||}{\textbf{NLoS channel model}} & \multirow{2}{*}{\textbf{$h_{\rm ABS}$}}\\  
 & $x_1$ & $x_2$ & $x_3$ & $x_4$ &$n$ & $\sigma_0$ & $\sigma_\infty$ & $h_0^\sigma$ & $n_0$ & $n_\infty$ & $h_0^n$ & $\sigma_0$ & $\sigma_\infty$ & $h_0^\sigma$& \\ \hline

\rowcolor{gray!15}
\multicolumn{16}{|c|}{\textbf{Manhattan Layout}} \\ \hline
\multirow{2}{*}{Suburban} &\multirow{2}{*}{ -5.776}&\multirow{2}{*}{13.96}&\multirow{2}{*}{-12.28}&\multirow{2}{*}{1.945}& \multirow{5}{*}{2.0} & \multirow{2}{*}{3.9} & \multirow{2}{*}{1.2} & \multirow{2}{*}{46}  & 4.2 & 2.65 & 16  & \multirow{2}{*}{14.6}   & \multirow{2}{*}{11.2}    & \multirow{2}{*}{15} & $<100$\\
  & &&&& &  &  &  & 2.5 & 2.91 & 172 &     &     & & 100--1000 \\
Urban   &-3.579& 9.018 &  -9.537& 2.799& & 5.3 & 2.2 & 46  & 4.62 & 2.75 & 49  & 14.9 & 12.8 & 610 &  \multirow{3}{*}{$<1000$}\\
Dense Urban & -3.274& 8.074& -8.839 & 3.342&  & 5.7 & 2.8 & 80  & 4.62 & 2.80 & 91  & 16.2    & 7.6    & 7& \\
High-rise & -4.008& 9.809  & -10.23  & 4.849 & & 5.5 & 3.3 & 327 & 4.55 & 2.55 & 319 & 11.5    & 17.5    & 40 & \\ \hline

\rowcolor{gray!15}
\multicolumn{16}{|c|}{\textbf{SRU Layout}} \\ \hline
\multirow{2}{*}{Suburban}  &  \multirow{2}{*}{-9.31}&\multirow{2}{*}{20.71}&\multirow{2}{*}{-16.64}&\multirow{2}{*}{2.78}& \multirow{5}{*}{2.0} & \multirow{2}{*}{3.4} & \multirow{2}{*}{1.2} & \multirow{2}{*}{58}  & 4.48 & 2.67 & 15  & \multirow{2}{*}{18.6}   & \multirow{2}{*}{11.5}    & \multirow{2}{*}{12} & $<100$\\
  & &&&& &  &  &  & 2.5 & 2.94 & 167 &     &     & & 100--1000 \\
Urban   &-4.933& 12.40&-12.83&4.049& & 4.5 & 2.2 & 44  & 4.90 & 2.79 & 42  & 18.5 & 13.9 & 58 &  \multirow{3}{*}{$<1000$}\\
Dense Urban &-4.253&11.13&-12.37&4.827&  & 4.3 & 2.5 & 72  & 4.86 & 2.85 & 74  & 18.4 & 15.3 & 171 & \\
High-rise &-13.16&37.89&-37.91&13.73&  & 4.2 & 3.1 & 325 & 4.81 & 2.64 & 252 & 16.5    & 18.3    & 8 & \\ \hline

\rowcolor{gray!15}
\multicolumn{16}{|c|}{\textbf{FUU Layout}} \\ \hline
\multirow{2}{*}{Suburban}  & \multirow{2}{*}{-16.54}&\multirow{2}{*}{30.55}&\multirow{2}{*}{-19.85}&\multirow{2}{*}{2.668}& \multirow{5}{*}{2.0} & \multirow{2}{*}{2.3} & \multirow{2}{*}{0.6} & \multirow{2}{*}{144}  & 4.04 & 2.72 & 19  & \multirow{2}{*}{24}   & \multirow{2}{*}{11.2}    & \multirow{2}{*}{16} & $<100$\\
  & &&&& &  &  &  & 2.54 & 2.95 & 124 &     &    &  & 100--1000 \\
Urban   &-6.686&16.24&-14.42&3.726& & 2.4 & 1.3 & 219  & 4.64 & 2.89 & 36  & 20.7 & 13.2 & 66 & \multirow{3}{*}{$<1000$} \\
Dense Urban &-2.772&8.748&-11.10&4.276&  & 2.8 & 1.7 & 133  & 4.76 & 2.94 & 66  & 19.4 & 12.8 & 328 & \\
High-rise &-6.721&18.93&-20.69&8.675&  & 3.2 & 2.3 & 260 & 4.72 & 2.74 & 238 & 21.7 & 16.7 & 182 & \\ \hline

\rowcolor{gray!15}
\multicolumn{16}{|c|}{\textbf{HEU Layout}} \\ \hline
\multirow{2}{*}{Suburban}  & \multirow{2}{*}{-11.49}&\multirow{2}{*}{23.93}&\multirow{2}{*}{-17.67}&\multirow{2}{*}{2.468}& \multirow{5}{*}{2.0} & \multirow{2}{*}{2.3} & \multirow{2}{*}{0.9} & \multirow{2}{*}{132}  & 4.19 & 2.78 & 12  & \multirow{2}{*}{22.7}   & \multirow{2}{*}{12.1}    & \multirow{2}{*}{11} & $<100$\\
  & &&&& &  &  &  & 2.41 & 2.94 & 76 &     &     &  & 100--1000 \\
Urban   &-7.536&17.33&-15.02&3.709& & 2.7 & 1.7 & 104  & 4.65 & 2.91 & 37  & 21.0 & 14.2 & 66 & \multirow{3}{*}{$<1000$} \\
Dense Urban &-5.589&14.63&-14.35&4.083&  & 3.2 & 2.0 & 76  & 4.62 & 2.95 & 57  & 20.6 & 14.6 & 182 & \\
High-rise &-7.308&21.05&-21.34&7.568&  & 3.2 & 2.5 & 306 & 4.58 & 2.80 & 195 & 22.3 & 16.1 & 356 & \\ \hline

\rowcolor{gray!15}
\multicolumn{16}{|c|}{\textbf{All layouts combined}} \\ \hline
\multirow{2}{*}{Suburban} &\multirow{2}{*}{ -12.5}&\multirow{2}{*}{24.25}&\multirow{2}{*}{-16.99}&\multirow{2}{*}{2.25}& \multirow{5}{*}{2.0} & \multirow{2}{*}{3.1} & \multirow{2}{*}{1.1} & \multirow{2}{*}{69}  & 4.28 & 2.70 & 14  & \multirow{2}{*}{19}   & \multirow{2}{*}{11.6}    & \multirow{2}{*}{12} & $<100$\\
  & &&&& &  &  &  & 2.53 & 2.93 & 138 &     &     &  & 100--1000 \\
Urban   & -4.66 & 11.42 & -11.45 &  3.369 & & 4.0 & 1.9 & 49  & 4.71 & 2.82 & 42  & 18.3 & 14.0 & 76 & \multirow{3}{*}{$<1000$} \\
Dense Urban & -3.922 & 9.727 & -10.19 & 3.826 & & 4.3 & 2.3 & 59  & 4.70 & 2.87 & 73  & 18.4 & 14.3 & 375 & \\
High-rise &  -3.929 & 9.645 & -10.26 & 5.137 & &4.1 & 2.8 & 220 & 4.64 & 2.68 & 253 & 16.6    & 18.5    & 7 & \\ \hline

\end{tabular}
\label{tab:params_full}
\end{table*}

\section{Model Validation \& Analysis}
\label{sec4}

\begin{figure*}[!t]
   \centering
    \begin{subfigure}[b]{0.24\linewidth}
    \includegraphics[width=\linewidth]{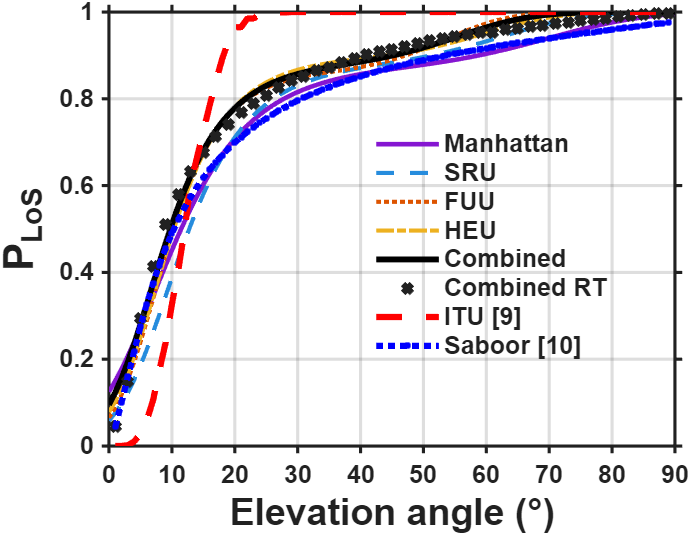}
    \caption{Suburban}
    \label{fig:SU_PloS}
  \end{subfigure}
  \centering
    \begin{subfigure}[b]{0.24\linewidth}
    \includegraphics[width=\linewidth]{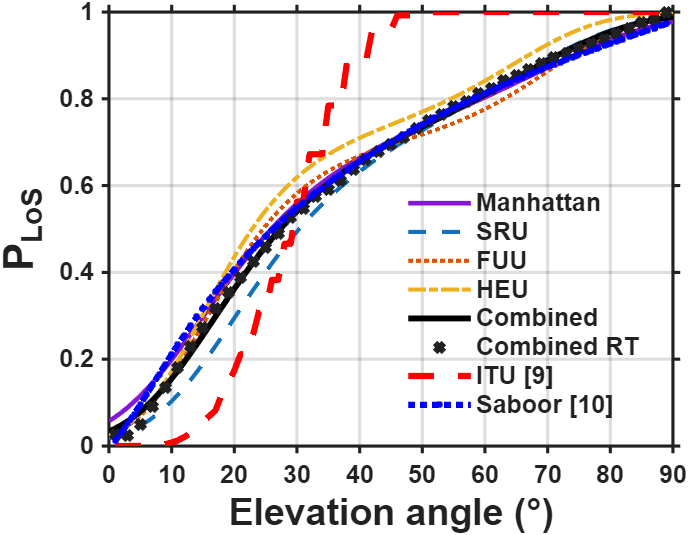}
    \caption{Urban}
    \label{fig:U_PloS}
  \end{subfigure}
  \centering
  \begin{subfigure}[b]{0.24\linewidth}
    \includegraphics[width=\linewidth]{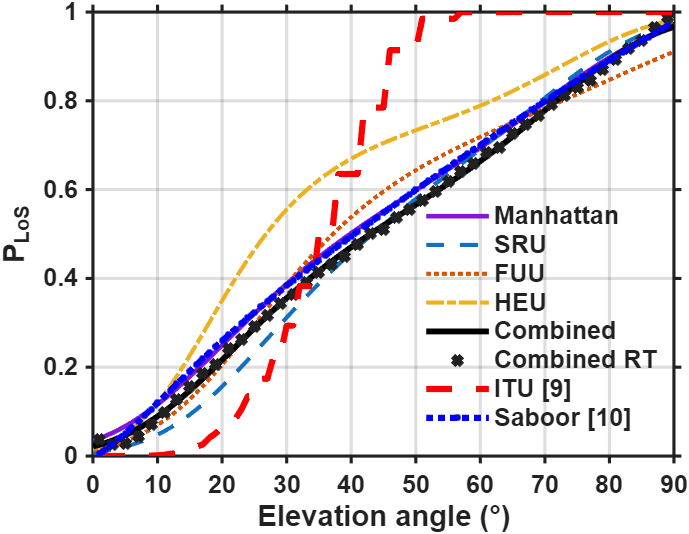}
    \caption{Dense Urban}
    \label{fig:DU_PloS}
  \end{subfigure}
  \centering
  \begin{subfigure}[b]{0.24\linewidth}
    \includegraphics[width=\linewidth]{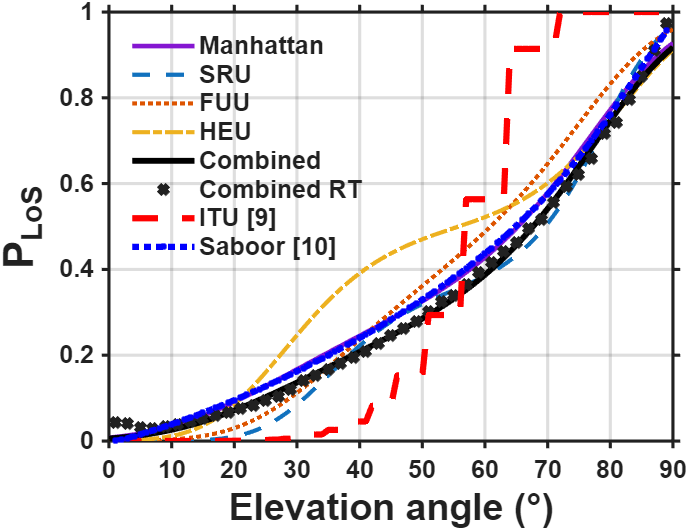}
    \caption{High-rise}
    \label{fig:HR_PloS}
  \end{subfigure}
  \caption{Empirical $P_{\text{LoS}}$ as a function of elevation angle for four standard environments, comparing individual layouts, combined RT data, and the unified combined model.} 
  \label{fig:PloS}
  \vspace{-1em}
\end{figure*}

Fig.~\ref{fig:PloS} illustrates the $P_{\text{LoS}}$ as a function of elevation angle. As expected, $P_{\text{LoS}}$ increases with ABS elevation. The results compare individual layouts, the combined RT dataset, and the unified model, along with the ITU model~\cite{ITU} and the geometry-based 3D model~\cite{Saboor2023plos}. The proposed model closely follows both the RT results and the 3D model across all environments, confirming its consistency and accuracy. In contrast, the ITU model shows noticeable deviations, particularly at moderate elevation angles, due to its simplified assumptions, as explained in~\cite{Saboor2023plos}.

Across environments, $P_{\text{LoS}}$ is highest in suburban and lowest in high-rise scenarios. Building height has a dominant influence on $P_{\text{LoS}}$, leading to relatively small differences between urban and dense urban cases. Overall, all layouts exhibit similar trends, indicating a limited impact of spatial arrangement for identical built-up parameters.

\begin{figure}[!t]

        \centering
        \includegraphics[width=1\linewidth]{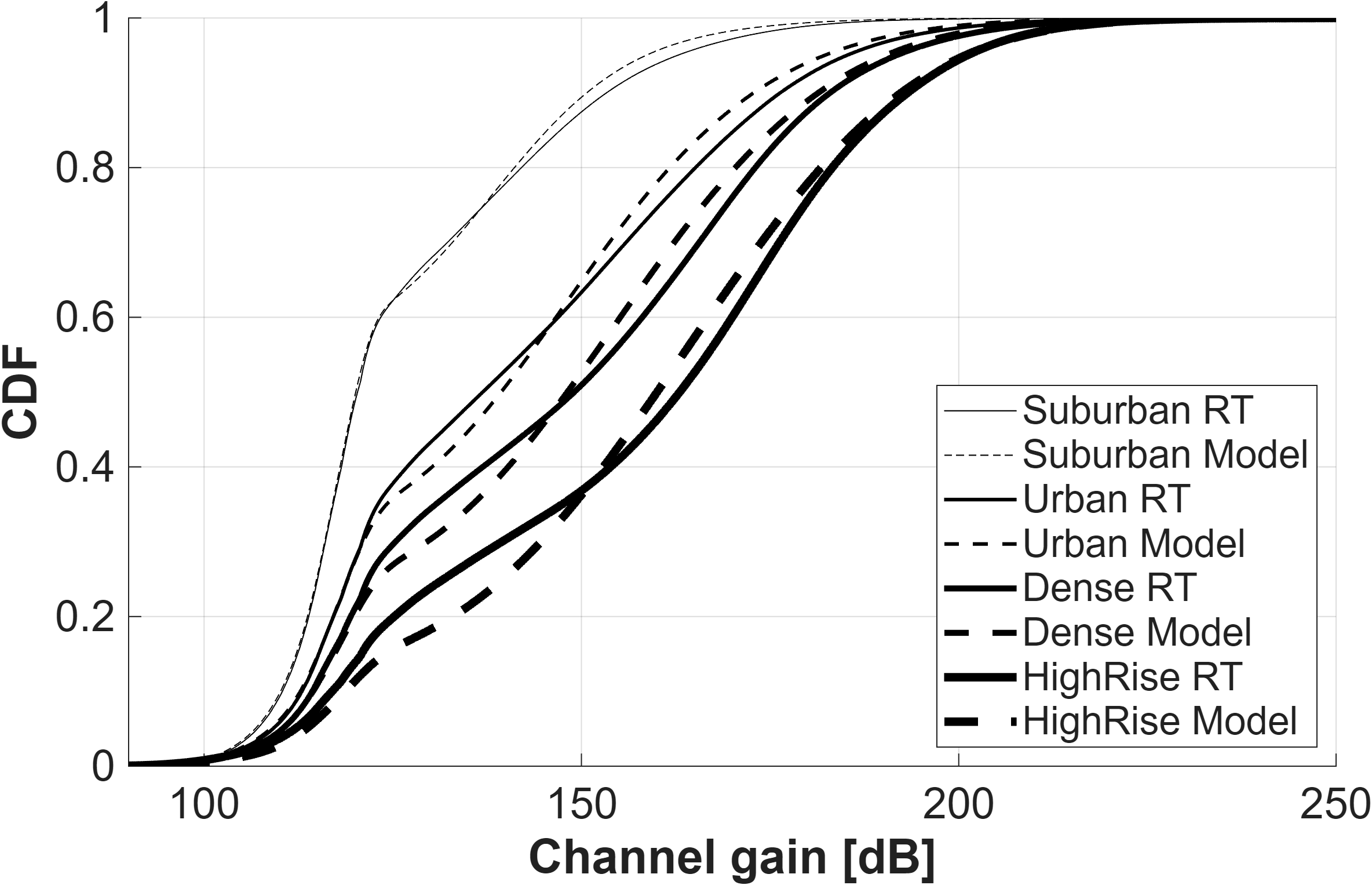}
        \caption{Comparison of modeled channel against RT data in the combined layouts case.}
    \label{fig:channel}
    \vspace{-.5cm}
\end{figure}

To further validate the proposed model, we compare its results with the full RT dataset by evaluating it over the same geometries (i.e., ABS and GU locations) used in the simulations. The comparison of channel realizations is shown in Fig.~\ref{fig:channel}, where visual inspection indicates good agreement between the RT simulation and our channel model in \eqref{eq:channel_model}. Table~\ref{tab:KL} quantifies this similarity using the Kullback–Leibler (KL) divergence~\cite{perez2008}, defined as
\begin{equation}
D_{\mathrm{KL}}(P \| Q) = \sum_{i} P(i)\,\log\left(\frac{P(i)}{Q(i)}\right),
\end{equation}
where $P$ and $Q$ denote the empirical (RT) and modeled distributions, respectively. The low divergence values confirm that the model accurately captures the statistical properties of the RT channel.

\begin{table}[t]
\centering
\footnotesize
\caption{KL divergence between RT data and channel model across environments and layouts}  
\begin{tabular}{|l|c|c|c|c|c|}
\hline
\rowcolor{gray!15}
\textbf{Environment} & \textbf{Manhattan} & \textbf{SRU} & \textbf{FUU} & \textbf{HEU} & \textbf{Combined} \\ \hline
{Suburban}   & 0.019 & 0.283 & 0.010 & 0.012 & 0.009 \\ \hline
{Urban}      & 0.020 & 0.041 & 0.022 & 0.027 & 0.041 \\ \hline
{Dense Urban}& 0.088 & 0.058 & 0.056 & 0.058 & 0.023 \\ \hline
{High-rise}  & 0.028 & 0.130 & 0.127 & 0.067 & 0.044 \\ \hline

\end{tabular}
\label{tab:KL}
\vspace{-.5cm}
\end{table}

\begin{figure*}[!t]
   \centering
    \begin{subfigure}[b]{0.24\linewidth}
    \includegraphics[width=\linewidth]{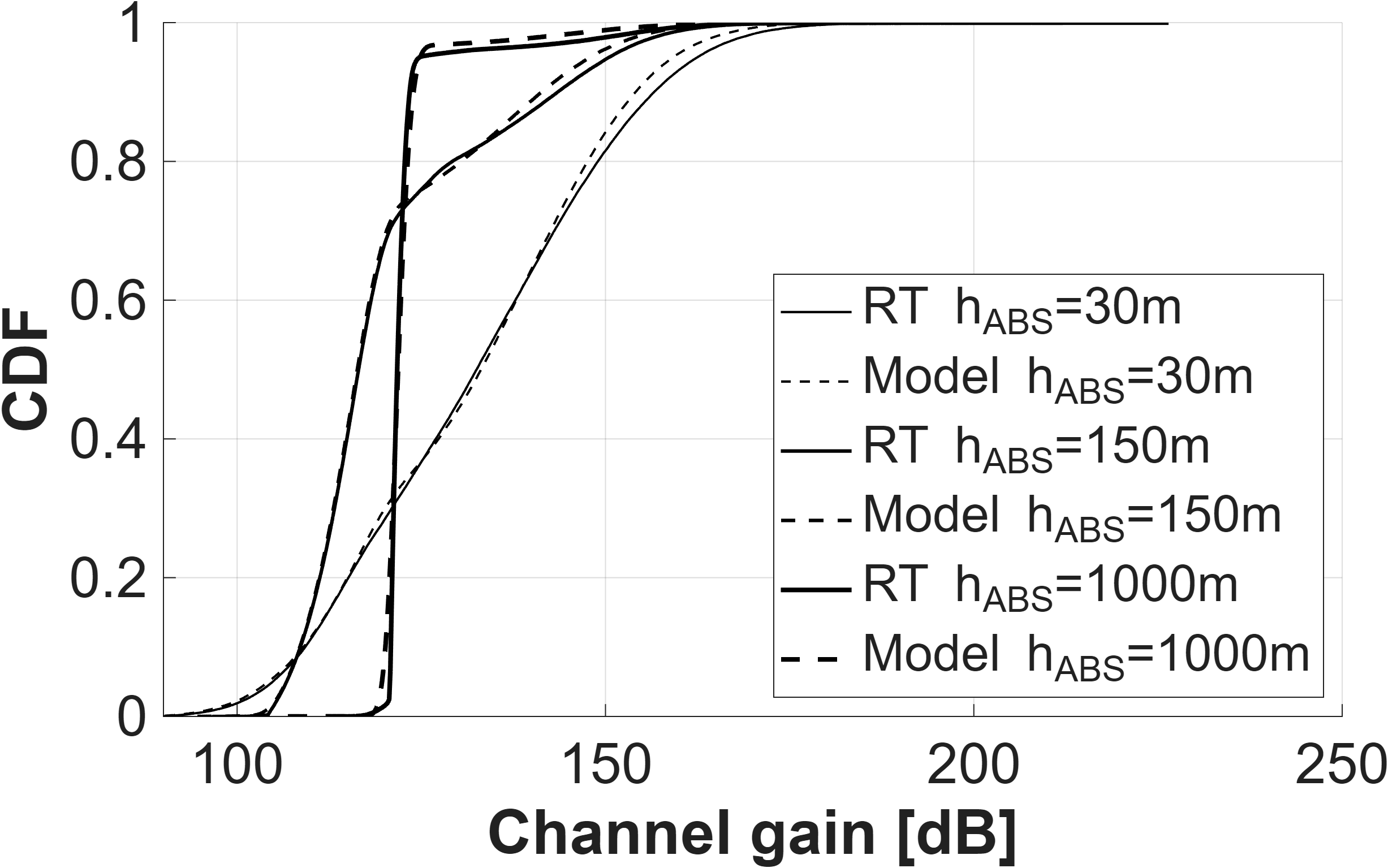}
    \caption{Suburban}
    \label{SU_CDF}
  \end{subfigure}
  \centering
    \begin{subfigure}[b]{0.24\linewidth}
    \includegraphics[width=\linewidth]{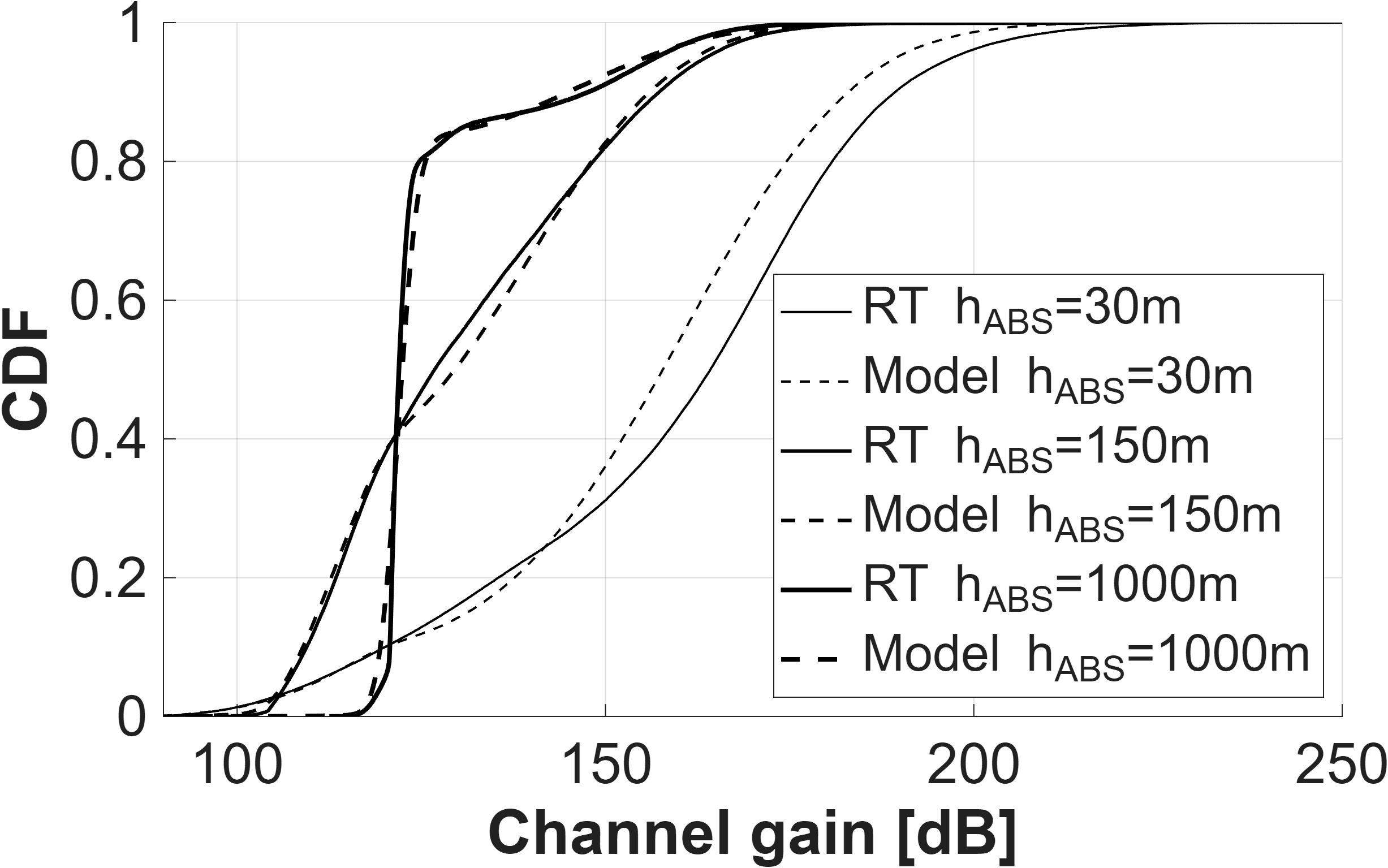}
    \caption{Urban}
    \label{U_CDF}
  \end{subfigure}
  \centering
  \begin{subfigure}[b]{0.24\linewidth}
    \includegraphics[width=\linewidth]{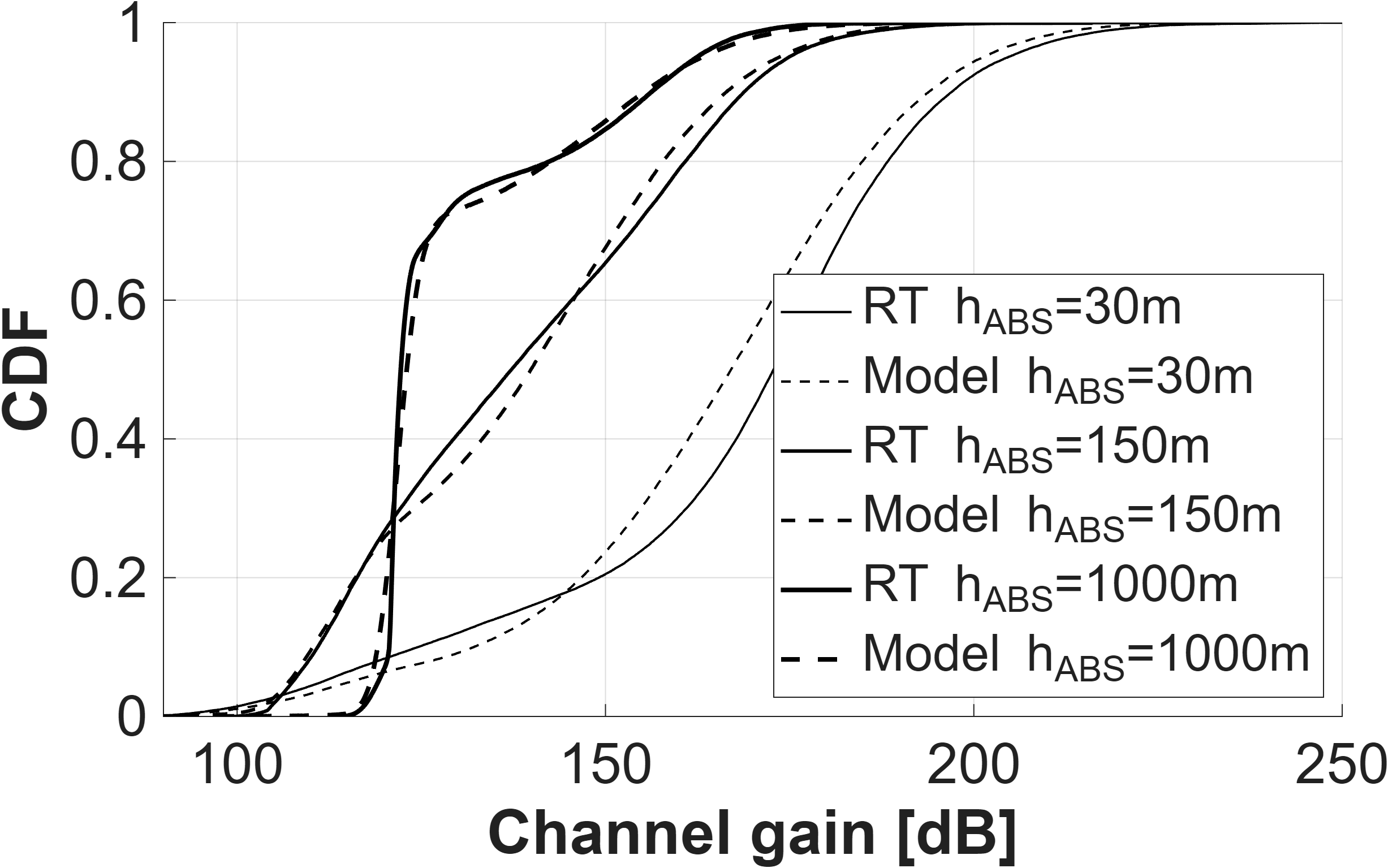}
    \caption{Dense Urban}
    \label{DU_CDF}
  \end{subfigure}
  \centering
  \begin{subfigure}[b]{0.24\linewidth}
    \includegraphics[width=\linewidth]{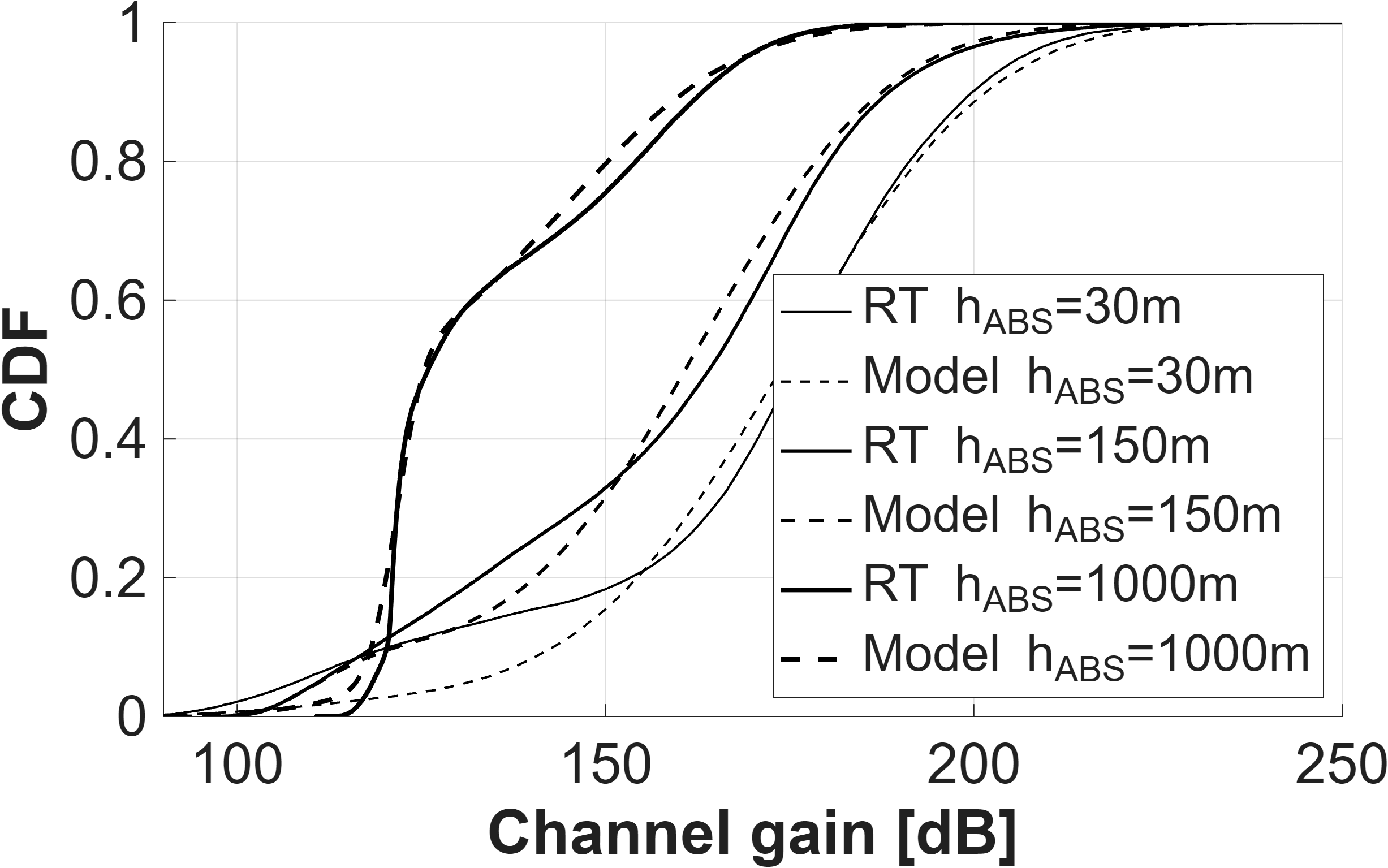}
    \caption{High-rise}
    \label{HR_CDF}
  \end{subfigure}
  \caption{Height-dependent channel model behavior compared against RT data for different environments. Each figure shows multiple ABS heights to verify both LSF and the complete channel model.} 
  \label{fig:channel_heights}
  \vspace{-1em}
\end{figure*}

Additionally, height-dependent behavior of the channel model is presented in Fig.~\ref{fig:channel_heights}, showing per-environment comparisons for multiple ABS altitudes. It can be observed that the model captures the behavior of the RT data at three characteristic altitudes: i) above the rooftop height (30 m); ii) maximum UAV flight height allowed in the EU (150 m); iii) most probable operational height of Advanced Air Mobility (i.e., flying taxi, 1000 m). Figs.\ref{fig:channel}~and~\ref{fig:channel_heights} validate the ability of the model to reproduce channel variations observed in the RT data.

Moreover, Figs.~\ref{fig:channel} and~\ref{fig:channel_heights} highlight the influence of the $P_{\text{LoS}}$ on large-scale channel attenuation. The NLoS PLE values in Table~\ref{tab:params_full} show moderate variation across environments, ranging from 4.28 to 4.71 at low altitudes and from 2.68 to 2.93 at higher ABS altitudes. However, the difference relative to the LoS case with $n=2$ is considerably larger. The higher $P_{\text{LoS}}$ observed in the suburban scenario (Fig.~\ref{fig:SU_PloS}) compared with the high-rise case (Fig.~\ref{fig:HR_PloS}) therefore explains the stronger channel gain and lower path loss in the suburban environment.

Lastly, the physical origin of the two-regime PLE behavior observed in the suburban environment (Fig.~\ref{fig:ple_model} and Fig.~\ref{fig:ple_suburban}) requires further investigation. A detailed inspection of the RT data provides preliminary evidence that diffraction is the dominant mechanism. When diffracted rays were excluded from the path loss computation, the PLE followed a single descending regime. These findings suggest that the sparse, low-rise building geometries can produce a relatively strong diffracted component at larger ABS altitudes, leading to coherent multipath interference effects similar to those in the classical two-ray propagation model. Similar behavior could potentially appear in other environments, but only at higher ABS altitudes (e.g., High Altitude Platforms or HAPs).

\section{Conclusion}
\label{sec5}
This paper examines how urban geometry, beyond built-up parameters, affects A2G propagation for UAV-based ABS deployments. Using MATLAB RT at 26 GHz across four layouts with identical built-up statistics, we show that NLoS PLE decreases toward $2.5$--$3$ with height, while LoS remains near $2$ and shadow fading reduces at higher altitudes. Although the overall trends are governed by $(\alpha,\beta,\gamma)$, spatial layout introduces a small PLE shift of about $\pm 0.2$, indicating that geometry has a measurable impact. The suburban environment exhibits two height regimes: a low-altitude region characterized by diffraction and shadowing, and a higher region (above 100 m) where reduced blockage and weak reflections result in a slight increase in PLE. Lastly, we proposed a unified height-dependent LSF model based on a four-parameter sigmoid for $P_{\text{LoS}}(\theta)$ and exponential expressions for PLE and shadow fading. The model closely matches RT statistics with low RMSE and KL divergence and generalizes well across all tested layouts. It provides a practical tool for ABS planning and 3D network design in complex urban areas. Future work will extend this framework to higher frequencies, mobility, and real measurement validation.

\section*{Acknowledgment}
This research is supported by the iSEE-6G project under the Horizon Europe Research and Innovation program with Grant Agreement No. 101139291 and by the Spanish Ministry of Science, Innovation and Universities MICIU/AEI/10.13039/501100011033 and the European Union NextGenerationEU/PRTR through the Ram\'{o}n y Cajal grant RYC2024-051003-1.

\vspace{.2cm}
\noindent\textbf{Declaration of Interest:} None. 

\vspace{.2cm}
\noindent\textbf{Content Generated by AI:} ChatGPT was used only for language editing, grammar checking, and minor assistance.  All text and code produced with AI were fully reviewed and verified by the authors, and all scientific ideas, analyses, and conclusions are entirely the authors’ own.

\balance
\bibliographystyle{IEEEtranN}  
\bibliography{ref}

\end{document}